# Effect of dc voltage pulsing on high-vacuum electrical breakdowns near Cu surfaces

Anton Saressalo,[1,*] Iaroslava Profatilova,[2,†] William L. Millar,[2,3] Andreas Kyritsakis,[1]
Sergio Calatroni,[2] Walter Wuensch,[2] and Flyura Djurabekova[1]

[1]*Helsinki Institute of Physics and Department of Physics, University of Helsinki,
P.O. Box 43 (Pietari Kalmin katu 2), 00014 Helsingin yliopisto, Finland*
[2]*CERN, European Organization for Nuclear Research, 1211 Geneva, Switzerland*
[3]*Cockcroft Institute, Lancaster University, Bailrigg, Lancaster LA1 4YW, United Kingdom*

Vacuum electrical breakdowns, also known as vacuum arcs, are a limiting factor in many devices that are based on application of high electric fields near their component surfaces. Understanding processes that lead to breakdown events may help mitigate their appearance and suggest ways for improving operational efficiency of power-consuming devices. Stability of surface performance at a given value of the electric field is affected by the conditioning state, i.e., how long the surface was exposed to this field. Hence, optimization of the surface conditioning procedure can significantly speed up the preparatory steps for high-voltage applications. In this article, we use pulsed dc systems to optimize the surface conditioning procedure of copper electrodes, focusing on the effects of voltage recovery after breakdowns and variable repetition rates as well as long waiting times between pulsing runs. Despite the differences in the experimental scales, ranging from $10^{-4}$ s between pulses up to pulsing breaks of $10^5$ s, the experiments show that the longer the idle time between the pulses, the more probable it is that the next pulse produces a breakdown. We also notice that secondary breakdowns, i.e., those which correlate with the previous ones, take place mainly during the voltage recovery stage. We link these events with deposition of residual atoms from vacuum on the electrode surfaces. Minimizing the number of pauses during the voltage recovery stage reduces power losses due to secondary breakdown events, improving efficiency of the surface conditioning.

## I. INTRODUCTION

Electrical vacuum arcing is a phenomenon connected to any device or component operating under high electric fields. If a device operates in air, it is not able to withstand high electric fields, since the dielectric strength of air does not exceed 3.0 MV/m [1]. To improve this common device-limiting factor, the air can be replaced by a specific gas with a higher dielectric strength—or simply by a vacuum with a dielectric strength that is ultimately high. However, even in vacuum, arcing is not avoidable. Its appearance limits the operation of diverse applications, including vacuum interrupters, satellites, medical devices, miniature x-ray sources, free electron lasers, and particle accelerators, such as fusion reactor beam injectors or elementary particle colliders [2–7].

Even though these vacuum devices have been used for more than a hundred years [8–13], the exact description of the vacuum arcing process is still under investigation. Open questions revolve around how the arc is initiated, how its location on the electrode surface is determined, and from where the neutral atoms required for plasma conduction are taken. Regardless of recent progress in answering these questions [14], linking atomic-scale models with experimental observations is still an object of experimental and computational research [15–17].

One of the major open questions is what triggers the vacuum arcing [18–21]. There are different hypotheses proposed to explain the phenomenon. For instance, surface impurities, dust, and other types of contamination are believed to be main factors triggering vacuum arcing near the metal surface [22]. However, the extent of this effect is not yet fully clear, since surface conditioning (exposure of the surface to a pulsed electric field for a long time) is believed to clean the surface from impurities (by detaching the particles from the surface or by burning them away in localized vacuum arcs) and, hence, to reduce the effect of these extrinsic factors drastically. However, there are indications that, after the surface has been conditioned, the arcing susceptibility still exhibits a dependence on the electrode material, revealing the intrinsic nature of conditioning [19,23–28]. Understanding to what extent both the extrinsic and intrinsic factors affect the vacuum arcing may help in focusing the efforts for mitigating this phenomenon.

*anton.saressalo@helsinki.fi
†iaroslava.profatilova@gmail.com

The phenomenon of vacuum arcing, or a vacuum electrical breakdown (BD), is especially crucial for the Compact Linear Collider (CLIC), a linear electron-positron collider proposed to be built at CERN, where high electric fields are required to make the accelerating length as short as possible. In the first stage, the particles are accelerated to energies up to 380 GeV over the course of around 5 km, leading to accelerating voltages of more than 75 MV/m. Copper has been selected as the material of the accelerating structures, which are essentially waveguides for electromagnetic radio-frequency (rf) pulses (11.9 GHz). The accelerating structures operate at room temperature in ultrahigh vacuum [10,29].

The rf structures are being investigated in the test facilities called rf test stands at CERN [30]. However, these extensive facilities require a large amount of resources to develop and operate. To tackle the BD problem purposely, a more compact system for generating BDs with direct-current (dc) voltage pulses has been designed and used at CERN and later also installed at the University of Helsinki.

In experiments where metal surfaces are exposed to high electric fields, it is a common practice to perform conditioning of the surface before the experiment to reach the highest conditioning state. Since numerous BD events will take place during the conditioning, the system needs a recovery procedure to return back to the pulsing mode after such an event. In a recent study [31], it was shown that the voltage recovery procedure influences the BD probability distribution function (PDF) over the number of pulses between two consecutive BDs. It was noticed that the probability was significantly higher for a BD to occur during the first pulse right after a stepwise voltage change during the recovery. It is not clear what may cause such an increase, since there are, in fact, two factors during the change of the ramping step: a change in the voltage and a 20-s pause needed to set a new value of the voltage.

Since the dc pulsed system produces pulses with higher frequency (up to 6 kHz) than the ones generated in the rf test stands at CERN (50–400 Hz), for compatibility of the conditioning results, it is also important to understand how the breakdown rate may depend on the pulsing frequency, also known as the pulsing repetition rate.

In this study, we performed various breakdown rate experiments on Cu electrodes. We compared the effects of different voltage recovery algorithms, pulsing repetition rates, and pauses of varied lengths between pulsing runs in order to understand how the electrode surfaces are cleaned during electric pulses and breakdowns.

## II. EXPERIMENTAL APPARATUS AND MEASUREMENT TYPES

The experiments were concluded with similar pulsed dc systems installed at the University of Helsinki and at CERN. The systems contain a power supply, a Marx generator [32], and a large electrode system (LES) combined with data acquisition and measurement electronics [31,33]. A schematic of the full system can be seen in Fig. 1.

Two cylindrical electrodes made out of copper [34] are placed inside the vacuum chamber of the LES with a distance of typically 40 or 60 $\mu$m and vacuum pressure below $1 \times 10^{-7}$ mbar when pulsing. The Marx generator is used for generating square dc pulses with voltages up to 6 kV (150 MV/m with a 40 $\mu$m gap and assuming $E = V/d$) and with a pulse width of typically 1 $\mu$s. The Marx generator also monitors the current during pulsing. When a peak in the current exceeds a threshold value, the device detects a breakdown and stops pulsing.

Another way of tracking the breakdowns is monitoring the vacuum pressure with an ion gauge. The readings show a pressure spike every time a breakdown occurs. Combining different methods of breakdown detection allows more accurate information on each event and makes it possible to find if some of the events are either falsely detected as BDs or real BDs not detected by the generator.

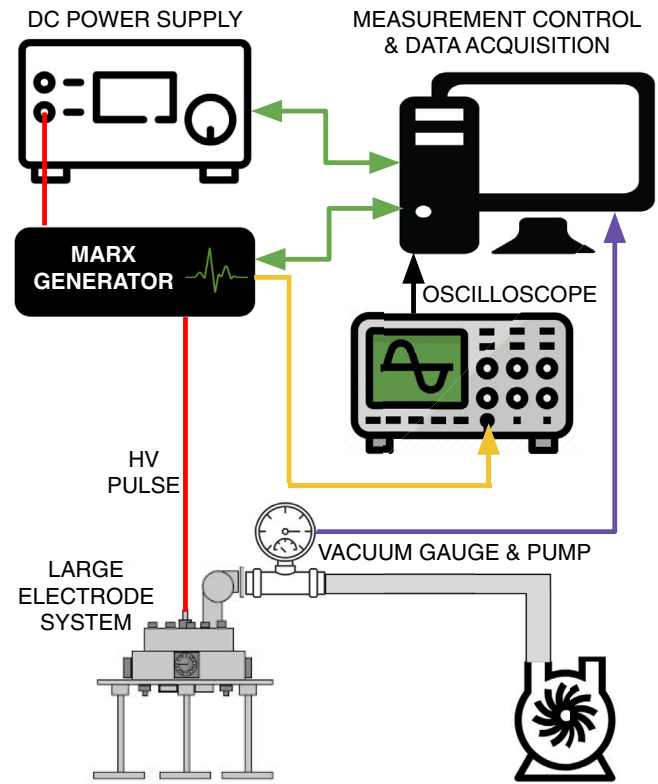

FIG. 1. A schematic of the pulsed dc system used in the measurements.

Other key parts of the measurement system include an oscilloscope for monitoring the voltage and current waveforms during pulsing and breakdowns as well as an ion gauge for measuring the vacuum pressure.

### A. Pulsing modes

A majority of the measurement runs were conducted using either of the two pulsing modes discussed below—or a combination of both.

In feedback mode, the target breakdown rate (BDR) is $10^{-5}$ BDs per pulses (bpp). This means that the pulsing is done in periods of typically 100 000 pulses (to match the target BDR). If a BD occurs during the period, the pulsing is immediately stopped, and the voltage is either decreased or kept constant, depending on the BD pulse number as explained in more detail in Ref. [33]. Otherwise, the voltage is increased. The maximum change after one period is typically set to $\pm 10$ V, which equals $\pm 0.16$ MV/m with a 60 $\mu$m gap. A pulsing run is defined as a combination of pulsing periods and breakdown events that occurred during one continuous experiment without additional pauses.

The feedback mode is used to condition the electrode surfaces. After being exposed to air—especially with pristine electrodes—they need to be conditioned to be able to operate at the highest possible electric field with a reasonable BDR. The conditioning is typically started with a low electric field, such as 10 MV/m. The feedback algorithm gradually increases the voltage over millions of pulses until the number of breakdowns starts increasing and the voltage level saturates, typically close to 100 MV/m [31,35,36]. The value varies slightly depending on the electrode type and the gap size.

In the other pulsing mode, which we call the flat mode, the voltage is kept constant during the whole measurement run, except for the voltage recovery after each BD. Usually, this mode is used only after the specimen has been conditioned and the flat mode voltage level is chosen close to the saturation value reached during the conditioning, so that the breakdown rate fluctuates close to the target value. However, sometimes choosing the correct level is difficult, and some conditioning effect is seen in the form of BDR fluctuations also during the flat mode.

### B. Voltage ramping scenarios

To recover the voltage after a breakdown, we apply a voltage ramp procedure; i.e., we increase the voltage gradually, ramping it from an initially low value to the targeted one ($\sim 5$ kV). Voltage ramp is performed after each breakdown in order to mitigate triggering of series of consequent breakdowns, which are known as secondary breakdowns (sBDs) [37]. These are seen to correlate with the preceding breakdowns spatially and temporally; i.e., each consequent event takes place in the vicinity of the preceding one after a relatively small number of pulses [31]. In contrast, the primary breakdowns (pBDs) are independent; they take place after a large number of pulses after the preceding breakdown at a random position on the material surface.

The voltage during the voltage ramp in the pulsed dc system is determined as

$$V_i = (V_{\text{target}} - V_{\text{start}})\left[1 - \exp\left(\frac{-P_i}{F \times P_{\text{step}}}\right)\right] + V_{\text{start}}, \quad (1)$$

where $V_i$ and $P_i$ are the voltage and the pulse number at the beginning of each ramping step $i$, respectively. $P_{\text{step}}$ is the number of pulses per each step, and $F$ determines the curvature. The shapes of different ramping modes obtained with different $F$ parameters can be seen in Fig. 2. As one can see, the $F$ parameters between 1 and 4 produce smooth curves asymptotically approaching the target value. With $F < 1$, the ramping voltage rises as a step function, while with $F > 10$, the voltage rises almost linearly. The starting voltage $V_{\text{start}}$ is determined by the ramping factor and is typically set to one-fifth of the target voltage $V_{\text{target}}$, i.e., the next voltage in the feedback mode or the set voltage of the flat mode. During the voltage ramp, there is always a pause of $\sim 20$ s before a change in the voltage value is applied. These pauses introduce additional idle time in the system before the pulsing is applied again.

Voltage ramp is applied only after a breakdown. After a pause due to other reasons, such as changing the pulsing period, the target voltage is applied directly without a ramp.

In our system, the voltage can be ramped stepwise or linearly. In the former case, the voltage is changed abruptly at the end of each step, during an unavoidable pause. In the latter, the voltage is gradually increased according to a given slope without pauses during the pulsing. The linear voltage ramp scenarios with different slopes were implemented to investigate whether a gradual increase in the voltage is more beneficial for the optimal voltage ramp scenario. In order to mimic the asymptotic behavior of the ramping procedure with $1 < F < 4$, we applied linear voltage ramp scenarios with multiple slopes. Changing the slope of the linear ramp also requires a 20-s pause; hence, we note that in all voltage ramp scenarios, except for the single-slope linear ramp, there was always an additional idle time needed for a voltage parameter to be changed. In the single-slope scenario, an idle time is not required even after the voltage ramp is complete.

In our previous studies [31,33,36], where a voltage ramp scenario with 20 steps and with curvature $F = 4$ was applied, we noticed an increase in BD probability right after the change in the voltage value at the beginning of the step, especially between 200 and 1000 pulses, where both the absolute voltage and the relative voltage change were significant.

Currently, we focus on different voltage ramp parameters to analyze the effect of this procedure on surface performance at high electric fields to optimize the surface conditioning. We compare the total BDR, the fraction of sBDs, and the average number of sBDs in a series triggered after a primary event, $\mu_{\text{sBD}}$.

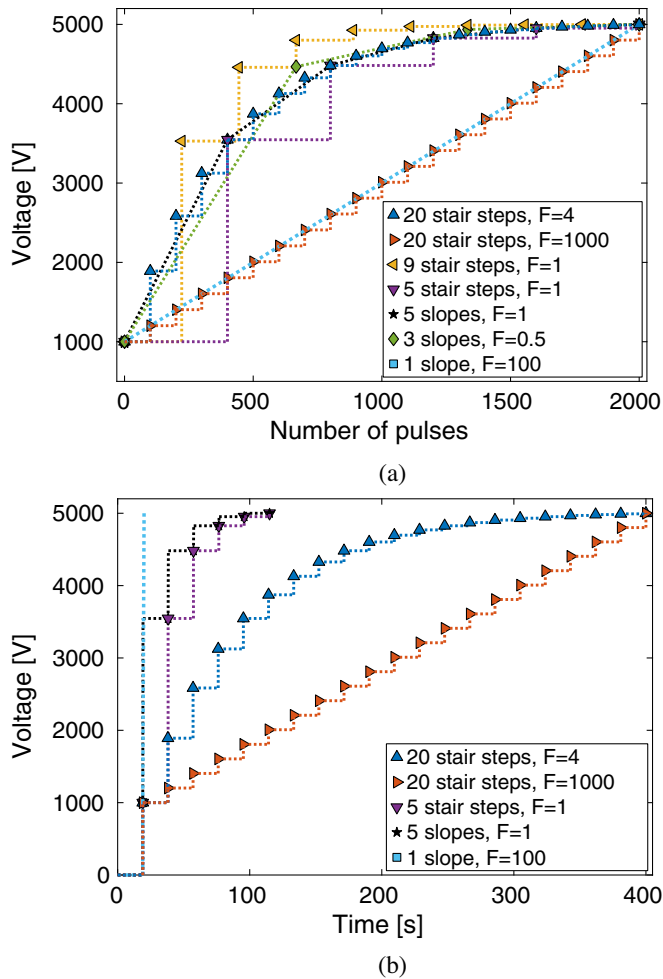

FIG. 2. (a) Comparison of five different ramping scenarios with voltage against number of pulses following Eq. (1) with $V_{\text{target}} = 5000$ V and $V_{\text{start}} = 1000$ V (ramping factor $\frac{1}{5}$). Each marker represents the voltage at the beginning of each ramping step or slope. A 20-s pause (idle time) in the system always precedes the pulse with the marker. (b) The selected scenarios with respect to elapsed time after the previous breakdown. In the beginning of each step, there is a fraction of a second of pulsing followed by nearly 20 s of idle time during which the voltage change is performed. The voltage increase in the slope scenarios is not instant but takes place during this fraction of a second, and, thus, the slope is not really visible in the graph.

The sBDs were determined by fitting the BD probability density function $[\rho_{\text{BD}}(n)]$ of the number of pulses $n$ between the two consecutive BD events by a two-term exponential model:

$$\rho_{\text{BD}}(n) = A\exp(-\alpha n) + B\exp(-\beta n) \qquad (2)$$

as introduced in Ref. [37]. The cross point of the two-exponential curves was used as the dividing line so that the BDs that occurred at a smaller number of pulses were defined as secondaries and the ones with a larger number of pulses as primaries. The exponential coefficients correspond to the breakdown rate of each regime, $\alpha$ to BDR of primaries, and $\beta$ to that of secondaries.

### C. Repetition rates of pulsing

The nominal repetition rate for CLIC energy stages is 50 Hz [11]. Nevertheless, the latest klystron-based X-band rf test facilities at CERN can operate at repetition rates of up to 400 Hz in order to reduce the time required to precondition the accelerating structures [38]. However, the effect of repetition rate on conditioning is not well understood, and breakdown rate measurements performed in the test stands typically require long time frames of the order of months, and, as a consequence, there is currently little literature on the subject.

High repetition rates available in the pulsed dc systems offer a unique opportunity to clarify this effect. Previously, a potential relationship between repetition rate and BDR was investigated experimentally in both the rf (25–200 Hz) and dc (10–1000 Hz) test stands. The results showed a small BDR increase at lower repetition rates; however, it was concluded that the observed difference was statistically insignificant and, hence, was suggested to be negligible [39]. Since a newly installed pulse source, such as a Marx generator, allows for a wider range of repetition rates in the pulsed dc system, the sensitivity of the BDs to the repetition rate can be measured with higher accuracy. The following steps were designed to perform the experiment: (i) Choose several values of repetition rates in the range from 10 to 6000 Hz in increasing order and apply high-voltage pulses using the flat mode. (ii) Choose two values of repetition rates and swap them several times to prevent electrode conditioning from masking the effect of the repetition rate changes. Reference frequencies of 100 Hz and 2 kHz were chosen and regularly compared between the measurements with different repetition rates. This also mitigated the effect of the BD clustering as the sBD series do not generally continue after a repetition rate change. (iii) In the last step, the so-called burst mode, the repetition rate is different for odd- and even-numbered pulses. This also means that the idle times before the odd-numbered pulses at 100 Hz are 2 orders of magnitude longer than those before the even-numbered pulses at 2 kHz. The pulse periods are then 10 and 0.5 ms, respectively. The burst mode is visualized in Fig. 3.

The measurements were calibrated and a voltage correction was made for all the repetition rates to adjust the voltage droop that varies between the repetition rates. Additional checks before each experiment were performed to ensure that the difference in the voltage pulse amplitude was no more than 0.05%.

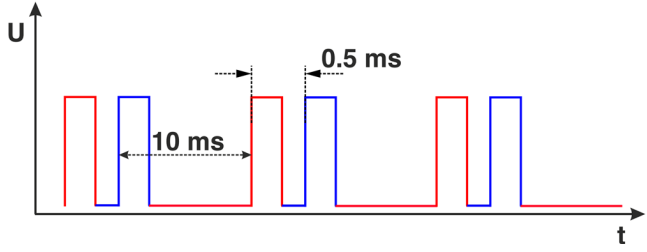

FIG. 3. Schematic of the burst mode, where pauses between odd- and even-numbered pulses (red and blue, respectively) are different. Note that the x axis is not linear, as the 1 μs pulse width has been exaggerated for visualization.

### D. Pauses between pulsing runs

In many rf test stand and pulsed dc system breakdown experiments, it has been qualitatively noticed that the breakdown probability is notably increased during the first pulses following a significant pause in operation. This effect has been noticed in experiments even when vacuum integrity was maintained throughout the pause, typically in the ultrahigh vacuum range. To date, however, this phenomenon has not been the subject of a purpose study.

In the present study, we analyzed this effect by carrying out a number of experiments with pauses ranging from 15 to $1 \times 10^5$ s and $4 \times 10^4$ to $2.5 \times 10^5$ s between pulsing runs with the dc and rf systems, respectively. High or ultrahigh vacuum was maintained during the pause. We quantify the effect of these pauses in both the pulsed dc system and the rf test stands by recording the number of breakdowns triggered in the system right after the pause.

In the dc case, the initial BDs after the random length pause were defined as the events that occur within the first second of pulsing (2000 pulses) with a constant voltage. Each pause was preceded by 50 000 pulses without a BD to ensure that the system was always in the same condition when the pause started. Hence, no voltage ramping was used after the pause. In the rf case, the system was run continuously for 2 h after the pause, and the probability for BDs to occur during this time was calculated as breakdowns per pulses.

## III. RESULTS

### A. Studies of different ramping scenarios

Copper electrodes with a 40 mm anode against a 60 mm cathode and a 40 μm gap were used in this experiment.

Seven different ramping scenarios and one without ramping were investigated in the flat mode at a voltage close to the saturation value, with ~1000 breakdowns and at least $10^7$ pulses in each run. Each flat mode run was preceded by a short feedback mode run in order to check that there were no drastic changes in the conditioning state of the electrodes which would affect the saturation voltage. All the ramping experiments were conducted with a repetition rate of 2 kHz.

The ramping scenarios are listed and visualized in Fig. 2. They include four stepwise voltage ramps and three cases where the voltage was ramped in slopes. In each of these cases, the ramping parameters were chosen to match the shape of the voltage ramp used in the previous experiments [31,33] (20 steps over 2000 pulses with $F = 4$). For comparison, we also performed an experiment without any ramping; i.e., the voltage was set to the target value immediately. In all scenarios, the electrostatic field reached at the end of the ramp was always between 126 and 128 MV/m. The results are listed in Table I and visualized in Fig. 4.

In the table, each voltage ramp scenario is described by the number and the method of the voltage changes, e.g., *20 steps* or *five slopes*, until the voltage reaches the target value. The $F$ parameter specifies the shape of the ramping curve—see Eq. (1). BDR is the breakdown rate measured in BDs per pulse.

We also show the fitting parameters $\alpha$ and $\beta$ [see Eq. (2)], which are essentially the BDRs of the primary and secondary events, respectively. The fit was done with an upper limit at 20 000 pulses and a lower limit at the number of pulses where the ramping voltage exceeds 95% of the target value, except the five-slope and three-slope scenarios, where the limits had to be manually adjusted. $N_{\text{cross}}$ in the table indicates the number of the pulses, at which the two exponentials describing sBDs and pBDs (see Sec. II B) intersect. All BDs, which were registered before $N_{\text{cross}}$, were defined as secondary ones. These are shown as the fraction of the total number of BDs in each run in percent ($N_{\text{sBD}}$). For the type with no ramping, the two-term exponential model was impossible to fit and the cross point was assumed at 1000 pulses, around the value where the PDF drops below $1 \times 10^{-5}$. The last column of the table shows the number of sBDs that took place after each pBD on average ($\mu_{\text{sBD}}$). All the uncertainties are calculated from the standard error of the mean.

TABLE I. Numerical comparison of the eight different voltage ramping scenarios, each described by the number of steps or slopes. Each was measured during a flat mode run with a repetition rate of 2 kHz, after a relaxation by conditioning. The $F$ parameter specifies the shape of the ramping curve, BDR is breakdowns per pulse, $\alpha$ and $\beta$ are fitting parameters of the two-exponential model, $N_{\text{cross}}$ is the number of pulses at which those exponentials intersect, $N_{\text{sBD}}$ is the fraction of all BDs that occur below $N_{\text{cross}}$, and $\mu_{\text{sBD}}$ is the mean number of consecutive BDs below $N_{\text{cross}}$. The uncertainties are calculated as the standard error of the mean. For the scenario with no ramping, the two-exponential fit was impossible, meaning that the values denoted with † are not fully comparable to the others.

| Ramp scenario | $F$ | BDR [bpp] | $\alpha$ | $\beta$ | $N_{\text{cross}}$ | $N_{\text{sBD}}$ [%] | $\mu_{\text{sBD}}$ |
|---|---|---|---|---|---|---|---|
| 20 steps | 4 | $1.78 \times 10^{-5}$ | $1.1 \times 10^{-4}$ | 0.002 | 2467 | $70 \pm 2$ | $3.3 \pm 0.1$ |
| 20 steps | 1000 | $7.68 \times 10^{-5}$ | $1.8 \times 10^{-4}$ | 0.010 | 2304 | $82 \pm 1$ | $4.3 \pm 0.2$ |
| Nine steps | 1 | $8.18 \times 10^{-5}$ | $1.8 \times 10^{-4}$ | 0.011 | 1205 | $88 \pm 1$ | $7.4 \pm 0.6$ |
| Five steps | 1 | $2.05 \times 10^{-5}$ | $5.3 \times 10^{-4}$ | 0.025 | 1385 | $84 \pm 2$ | $5.3 \pm 0.4$ |
| Five slopes | 1 | $3.37 \times 10^{-5}$ | $2.8 \times 10^{-4}$ | 0.010 | 1357 | $90 \pm 1$ | $9.8 \pm 0.8$ |
| Three slopes | 0.5 | $2.98 \times 10^{-5}$ | $5.2 \times 10^{-4}$ | 0.013 | 899 | $95 \pm 1$ | $16.7 \pm 0.9$ |
| One slope | 100 | $5.86 \times 10^{-6}$ | $1.1 \times 10^{-4}$ | 0.003 | 3666 | $67 \pm 1$ | $3.1 \pm 0.1$ |
| No ramp | ⋯ | $9.60 \times 10^{-5}$ | ⋯† | ⋯† | 1000† | $94 \pm 2$† | $15.3 \pm 1.3$† |

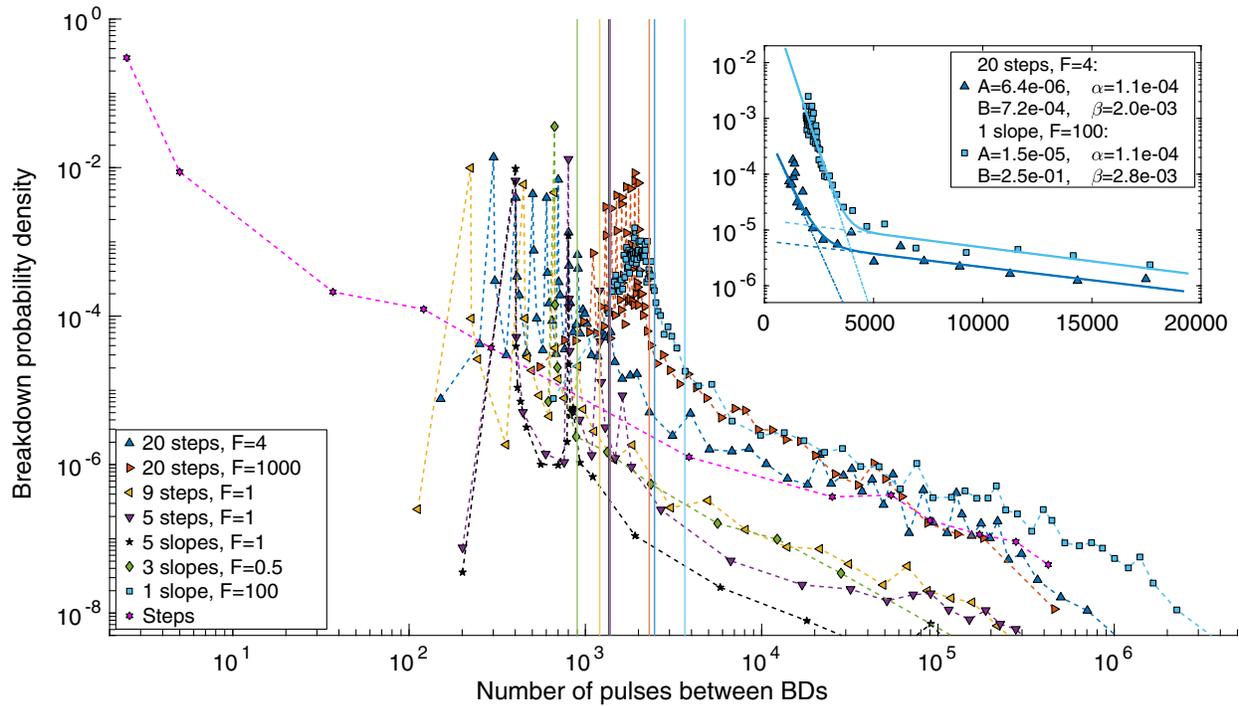

FIG. 4. Probability for a breakdown to occur at each number of pulses between breakdowns for each of the ramping scenarios. The vertical lines indicate the cross points of the two-exponential fit for each run. A visualization of the fit for the data points of 20 steps, $F = 4$ and 1 slope, and $F = 100$ is shown in the inset. There, the legend also shows the fit parameters.

The lowest BDR and the lowest sBD fraction can already point out the best ramping scenario. However, we note that these values, which are found for the ramp with 20 steps ($F = 4$) and the one with a single slope, are also strengthened by the lowest rates of the primary events $\alpha = 1.1 \times 10^{-4}$ and the smallest mean number of sBDs per a primary one, $\mu_{\text{sBD}} < 3.5$.

The ramping performance was also studied by plotting the PDF for a BD to occur after a given number of pulses after a previous breakdown, which is shown in Fig. 4. The figure also includes an inset showing the two-term exponential fit for the best two runs.

In this graph, one can clearly see a sawtooth behavior which correlates strongly with the ramping steps in each experiment, which was also observed in Ref. [31]. Each subsequent peak appears after the number of pulses that correspond to the length of a step or a slope. Since the change in the voltage (or the slope of the voltage increase) requires an additional idle time of about 20 s in the system, the combination of these two factors, i.e., increased voltage and the extra idle time before a new voltage step, leads to a higher probability of a breakdown during the first pulse after the pause.

The vertical lines in Fig. 4 show the cross points of the two-term exponential fits. The colors of these lines match the color of the markers chosen for each ramping run. It is clear that all of these lines are close to one another and are about 2000 pulses, which was the number of pulses used in all ramping procedures.

Since the voltage ramp always follows a postbreakdown pause, it is difficult to separate the effect of surface modifications caused by the preceding BD event and a possible effect of residual deposition during the pause time. The data showing the experiments with no ramping and with a single-slope ramping cast some light on this issue. We see that restoring the same voltage value immediately after the post-BD pause (no ramping) results in the highest BDR, showing practically a single-exponent behavior of $\rho_{\text{BD}}(n)$ in Fig. 4. Meanwhile, the linear increase in the voltage after a BD even with a postbreakdown pause resulted in the least secondary breakdowns during the ramping. This observation suggests that some relaxation effects take place on the surface during the gradual increase of surface charge.

We note that all the major ramping peaks appear before the cross points, hence concluding that almost all of the sBDs occurred during the ramping period. This also means that the sBDs barely separate from one another by more than 2000 pulses, which is the duration of the ramping procedure. Clearly, the changes taking place during the ramping affect the probability of triggering subsequent breakdowns, while at the target voltage, where all electric parameters are kept unchanged, the sBDs practically do not take place. In some runs, however, the cross point was found at a number of pulses greater than 2000, which means that, in these cases, the changes that took place during the ramping period still affect the surface behavior shortly after the pulsing at the target voltage has begun.

The breakdown probability was also analyzed as a function of the voltage, as shown in Fig. 5. Here we see that, during the stepwise ramping, breakdowns start taking place already at lower voltages, whereas the linear increase in the voltage without pausing leads to an increased BD probability at higher voltages and typically only after a slope change, i.e., a small pause in the pulsing. The type with no ramping was not evaluated, as all the breakdowns occurred at the target voltage.

### B. Experiments on pulsing repetition rates

Since we observe a clear correlation between the breakdown probability and the pause duration between the pulsing runs, we now turn our attention to the analysis on the effect of the repetition rate which we applied in different orders. All the repetition rate experiments were concluded using 40 mm Cu electrodes separated by a 60 $\mu$m gap.

#### 1. Increasing order of repetition rates

The pulsing was done with repetition rates ranging from 10 to 6000 Hz. The repetition rate was changed to the next one after every 100 BDs. The usual flat mode algorithm was used during each such step. The electric field was chosen to keep the BDR between $10^{-4}$ and $10^{-7}$ bpp. The results are shown in Fig. 6.

Figure 6(a) shows that the BDR (bpp) decreases as the repetition rate increases. With the same electric field, the BDR at 10 Hz is $7.3 \times 10^{-4}$ bpp, while at 6 kHz it is $6.6 \times 10^{-6}$ bpp, which is by 2 orders of magnitude lower. The difference is observable but less remarkable when the BDR is expressed as breakdowns per second (bps), as shown in the same figure. In the latter, we see a fivefold increase in the BDR from the lowest to the highest repetition rate (from $7.3 \times 10^{-3}$ to $3.97 \times 10^{-2}$ bps). The increase in the number of BDs per second is expected, since the idle time between the pulses decreases with an increase in the repetition rate. However, this difference in the idle time between the pulses in both regimes is much greater (600 times) compared to the observed increase in BDR ($\sim$5 times). Hence, the two graphs presented in Fig. 6(a) corroborate one another in spite of the difference in the measured rates.

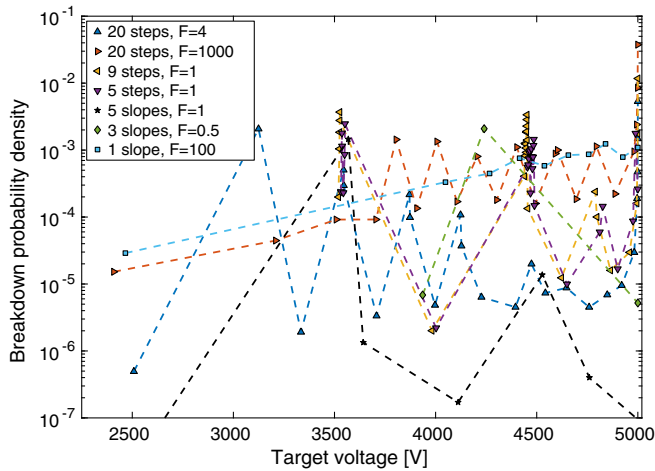

FIG. 5. Probability for a breakdown at a given ramping voltage. The x-axis values have been scaled so that $V_{\text{target}} = 5000$ V in each measurement.

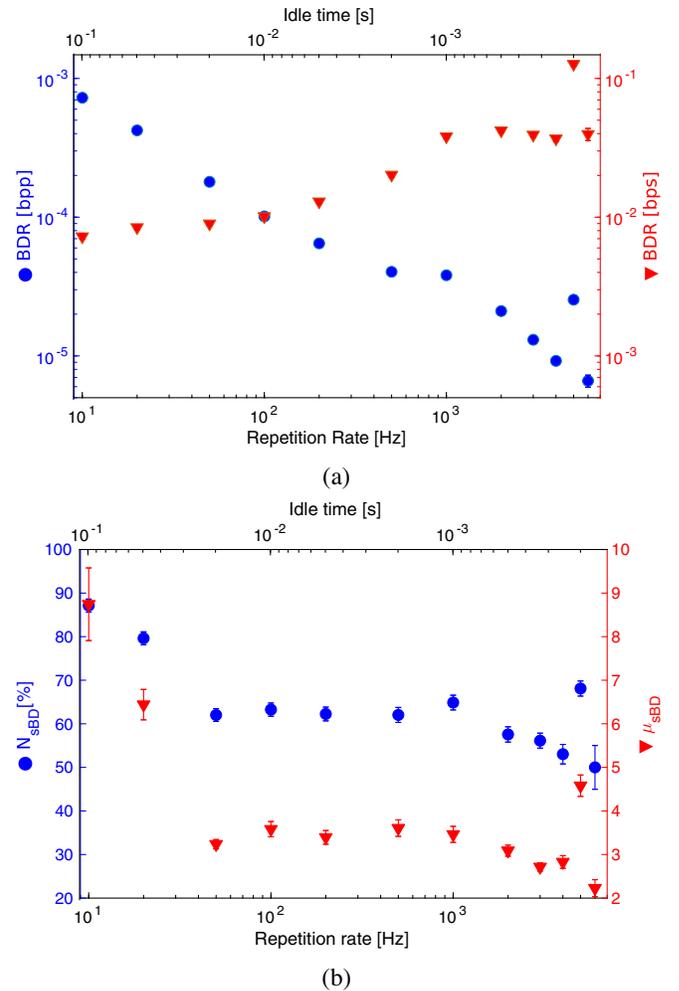

FIG. 6. BD experiments with variable repetition rates. (a) shows the BDR as BDs per pulse and as BDs per second for each repetition rate. (b) shows the fraction of secondary BDs $N_{\text{sBD}}$ and the mean number of secondary BDs after a primary one, $\mu_{\text{sBD}}$, for each repetition rate. In each graph, the uncertainties are estimated from the standard error of the mean, though many of the error bars are too small to be visible.

In order to exclude a possible effect of conditioning, which may explain the decrease of the BDR (bpp), we performed the same experiment in descending order of the repetition rates. The results presented in the Appendix A show a very similar trend as observed in Fig. 6(a), which allows us to conclude that the reduction of the BDR in bpp is mainly due to the decrease of the idle time between the pulses.

Since sBDs show a correlation with the preceding events, we plot the percentage of these events ($N_{\text{sBD}}$) separately in Fig. 6(b) along with the mean number of sBDs after a pBD ($\mu_{\text{sBD}}$), as a function of the repetition rate. Although the dependence is not as monotonic as in Fig. 6(a), the graphs clearly show that the values of both $N_{\text{sBD}}$ and $\mu_{\text{sBD}}$ are higher at lower repetition rates and decrease strongly at the higher rates.

#### 2. Swap repetition rates

During the pulsing experiments, the electrode surfaces are continuously conditioned, and this may affect the BDR measurements at different repetition rates, confusing the possible conclusions. To avoid the effect of change in the surface conditioning state between the measurements from screening the results, we applied a mode where the repetition rates were swapped between 100 Hz and 2 kHz after every three consecutive BDs that occurred at the target voltage. Figure 7 shows the cumulative number of BDs vs the number of pulses in the repetition rate swap regime (solid line). For comparison, we also show the same value accumulated during the 100 Hz repetition rate (red dash-dotted line) and that accumulated during the 2 kHz repetition rate (blue dash-dotted line).

We see that the BDR (in bpp) is approximately twice as high with the lower repetition rate, while it is practically the same as for the higher repetition rate in the experiments with the swapped repetition rates. It is clear that the system was running for a longer number of pulses when the repetition rate was high (blue segments) and started practically immediately breaking down, when the repetition rate was switched to the lower value (red segments in the shape of steps).

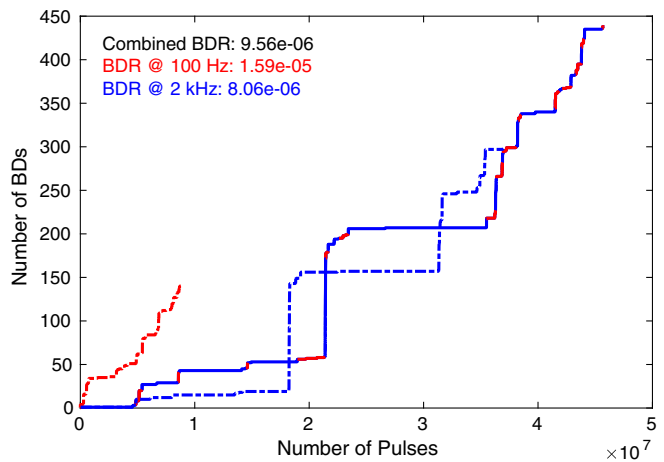

FIG. 7. Evolution of the cumulative number of BDs against the number of high-voltage dc pulses with the swapping of repetition rates. The solid line shows the result from the whole experiment, while the blue and red dash-dotted lines show the results for the repetition rates of 2 kHz and 100 Hz, respectively.

#### 3. Burst mode

An additional burst mode was implemented in the hardware of the generator in order to further study the effect of the repetition rates as described in Sec. II C. The measurements were performed with electric fields between 61 and 62 MV/m.

For the analysis, the breakdowns that occurred at the odd and the even pulses were separated. In this case, the pause before an odd-numbered pulse was 10 ms, and the pause before an even-numbered pulse was 0.5 ms. The results presented in Table II show again that the BDR is higher for the events with a longer pause before the pulse.

### C. Pause between pulsing

The pause between measurements with the pulsed dc system was measured using the feedback mode and by implementing a randomly selected pause between 15 and 100 000 s ($\sim$28 h) in length, each following a 50 000 pulsing period without a BD. The pause lengths were grouped into eight bins, and the BD probability was estimated from the fraction of cases that lead to a BD within the first second of pulsing (2000 pulses) after the pause. It is also important to note that no voltage ramping was used after the pause, as there was no preceding BD.

TABLE II. The results for the test with burst mode. $R_{\text{BDR}}$ shows the BDR ratio with the lower repetition rate divided by that of the higher repetition rate. The uncertainties were estimated from the standard error of the mean and the propagation of error.

| Pause [ms] | BDs | BDR [bpp] | $R_{\text{BDR}}$ |
|---|---|---|---|
| 10 | 370 | $(1.64 \pm 0.09) \times 10^{-5}$ | $1.58 \pm 0.13$ |
| 0.5 | 281 | $(1.04 \pm 0.06) \times 10^{-5}$ | |

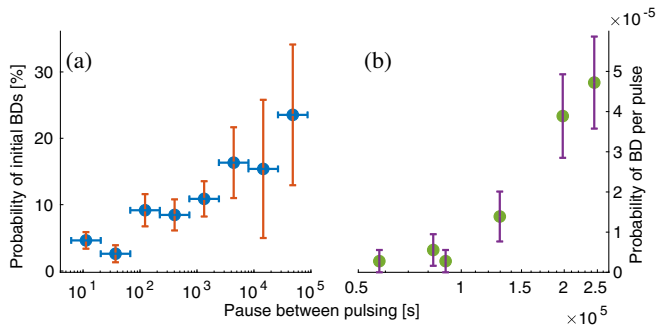

FIG. 8. (a) BD probability within the first second of running (2000 pulses) after a predetermined pause between pulsing runs, not preceded by a BD, with a pulsed dc system. The horizontal blue error bars indicate the range of the pause length values where the data were averaged. The vertical red error bars, again, indicate the uncertainty as the standard error of the mean. (b) The probability of BD, i.e., the BDR during the first 2 h of operation after a long pause in pulsing with a rf test stand. The violet vertical error bars indicate the uncertainty as the square root of the number of BD events.

For the rf test stand, the effect of the pause was estimated by presenting the number of BDs that occurred within the first 2 h of operation at 50 Hz (360 000 pulses) following a long pause in operation ranging from 11 to 68 h. With the rf test stand, the vacuum level is typically maintained at around $1 \times 10^{-10}$ mbar during the pauses, i.e., 2–3 orders of magnitude lower than in the pulsed dc systems.

The results of both tests are presented in Fig. 8 and both show that the BD probability increases with the length of the pause. This result indicates that there are processes, which take place during the pause and consequently affect the probability of BDs after the pause, when the electric field is restored at the surface. A more detailed analysis of the data presented in Fig. 8(a) can be found in Appendix B. The increasing trend is still clearly visible there in spite of the large error bars measured for the last three data points.

## IV. DISCUSSION

In the present study, we clearly see that the idle times between the pulsing runs and during the pulsing itself have a great impact on the BD probability. Several independent measurements show the same result: The longer the idle time between the pulses, the higher the BDR measured in bpp becomes.

Analysis on the different voltage ramp scenarios can shed some light on this effect. First of all, we see that both the shape of the ramping curve and the pauses between the ramping steps or slopes play important roles. While ramping the voltage stepwise, two changes appear between the steps: The voltage level increases, and the system pauses for 20 s to set the new voltage value. In the multislope voltage ramp, the voltage changes gradually; however, the system pauses for the same 20 s to set a new slope of the voltage ramp. Hence, by keeping one of the parameters, for instance, the number of pauses, intact, we are able to separate the effect of pauses and the voltage change on the BD probability.

Assuming that vacuum residuals could explain the increased BD probability in the system during the voltage ramp, one could suggest reducing the number of possible pauses to reduce the number of BDs and, hence, improve the efficiency of the surface conditioning. However, the results presented in Table I show that the nine and five pauses in the nine-step and five-step and slope scenarios resulted in higher BDR and fraction of sBDs as well as in longer BD series compared to the 20-step voltage ramps with both curvatures $F = 4$ and $F = 1000$. Moreover, the voltage ramp with three slopes, which required only three pauses, resulted in the worst result. Almost all the BDs were counted as secondaries, and the average number of sBDs in the series after a pBD is one of the largest. This indicates that the pauses (vacuum residuals) alone cannot explain the higher activity of the surface during the voltage ramp. The step height for the voltage ramp plays an important role as well.

As we see in Fig. 2(a), both nine- and five-step voltage ramps bring the voltage to 70% (and the three slopes to 90%) of the target value already at end of the very first step. Such high-voltage values following after the mandatory pause make these scenarios very similar to that with no ramping at all. Hence, we conclude that, although intuitively the large steps at low voltages are rather reasonable, the experiments show that the voltage should not change dramatically during the voltage ramp. In the following, we will obtain a deeper insight of the effect of the voltage change in the voltage ramps where the number of pauses was the same.

In Fig. 9, we enlarge the data presented in Fig. 4 to analyze the behavior of $\rho_{BD}$ for the sBDs, i.e., BDs with the enhanced probability that take place before the cross point of the two exponents. We selected four scenarios that can be grouped in pairs. The pairs are defined by the same number of pauses during the voltage ramp, either five or 20. However, within the pair, the voltage ramp was done differently. In the pair with five pauses, the voltage was ramped either in steps (five-step ramp) or in slopes (five-slope ramp). Both the height of the steps and the value of the slopes were modified during the ramp to follow the exponential shape of the ramp applied in Ref. [31].

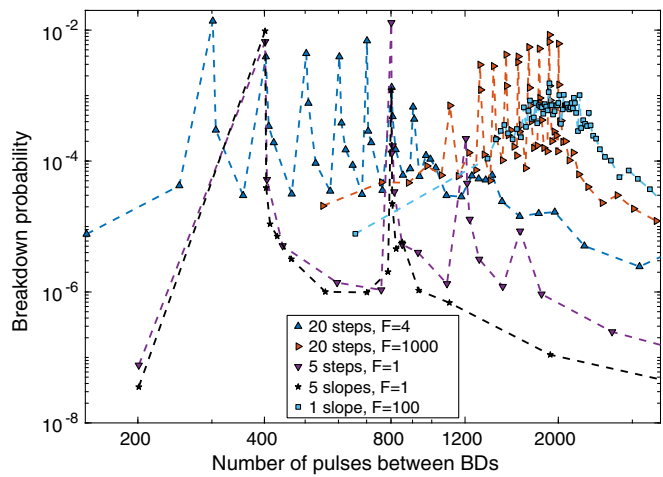

FIG. 9. Enlargement of Fig. 4 showing the breakdown probability PDF against the number of pulses between breakdowns for the scenarios with 20 stairs, five stairs, five slopes, and one slope.

In the pair with 20 pauses, the voltage was ramped in steps in both scenarios. The difference was in the height of the steps. In the ramp with $F = 4$, the step height varied similarly as it was done in Ref. [31], while in the ramp with $F = 1000$, the step height was kept constant, so that the overall ramping shape was linear.

Comparison of the curves for the five-pause pair in Fig. 9 reveals a striking similarity: Both curves have peaks at the same number of pulses between the consecutive BDs, at 400 and 800 pulses. Since, a single step before changing the voltage value or the slope consists of 400 pulses, it is clear that the first two pauses create conditions favorable for a BD to occur in the first pulse after the pause. In Fig. 2, we see that the voltage values right after the pauses in five-step and five-slope scenarios are the same. It is clear that the BD probability during the first pulse after a pause is increased due to an increased value of the voltage and due to exposure of the surface to the residual deposition during the pause. Although the system sees the increased value of the voltage already before the pause in the slopewise ramp, a BD is more probable to take place after the system paused. It is evident that the cleaning of the surface takes place even without BDs; otherwise, the BD events would be nearly equally probable in the slopewise ramping before and after the pause.

The second peak can be explained in a very similar manner as the first one. However, the further increase in the voltage values during the system pausing shows different behavior for both voltage ramps. The stepwise ramp exhibits two more peaks at 1200 pulses and at 1600 pulses, while the slopewise ramp does not show increased probabilities related to the pauses in the system. It is clear that the smoother change in the voltage of the slope provides gentler cleaning of the surface close to the target voltage value, compared to the stepwise changes, even if the system still pauses to change the slope. The lesser number of peaks in the $\rho_{\text{BD}}(n)$ for the five-slope voltage ramp does not, however, result in better overall performance. The data in Table I indicate that the stepwise voltage ramp gives a slightly smaller fraction of sBDs and a lower total BDR and can, hence, be considered as a more optimal scenario for the voltage ramp. This is due to higher intensity of the BD in the first pulse after the very first pause during the ramping. It appears that the pulsing with the steep linear voltage increase activates more spots for subsequent BDs ($\mu_{\text{sBD}}$ for the five-slope ramp is almost twice as high as for the five-step one) in the vicinity of the preceding ones than the pulsing at the same voltage.

The $\rho_{\text{BD}}(n)$ functions for both 20-step scenarios do not have identical peaks, although the number of pauses in both ramping scenarios is the same. The large steps in the voltage ramp in the beginning of the ramping procedure result in an increased BD probability for the ramp with $F = 4$. We do not register any breakdowns in the ramp with $F = 1000$ for the first six steps; however, we observe a strong increase of the BD probability with the well-pronounced peaks during the first pulse after the pauses at higher voltage steps. Moreover, the peaks are growing in height, illustrating that the voltage increase should be slower when approaching the voltage target value.

Despite the higher BD probability at the beginning of the voltage ramp, we see that, overall, the conditioning achieved in the voltage ramp with curvature $F = 4$ is more optimal, as it results in a lower total BDR and a lower fraction of the sBDs in the pulsing run. The BDs triggered at the voltage values closer to the target value are more intense and may result in a larger number of nuclei for the subsequent BDs. However, the difference in performance of the voltage ramps with 20 steps is not as dramatic as for the ramps with five steps and five slopes.

We also note that not only the number of steps but also the shape of the ramping scenario may play important role on the conditioning process. We notice that the 20-step scenario with $F = 4$ (exponential ramp) results in a gentler conditioning procedure than $F = 1000$ (almost linear ramp). This is explained by the greater number of steps spent in the latter scenario in the regime of low-voltage values, which produce very small initial breakdowns or do not trigger them at all. Moreover, the voltage increment per step with $F = 1000$ is much smaller in the early pulses, which may also affect the lower activation of initial breakdowns in this scenario.

Comparing the results obtained with the ramping scenarios containing different numbers of steps, we conclude that the pulsing at lower-voltage values is essential for cleaning the surface. If the surface was exposed to a sufficient number of lower-voltage pulses, the increased BD probability at higher-voltage values is less detrimental for the surface conditioning than in the reverse case (less low-voltage pulsing but reduced BD probability at the higher voltages). This conclusion is strengthened by the results shown in Fig. 9 for one slope. In this scenario, no pauses were allowed in the system, and the voltage was slowly ramped from the initial to the target value after a BD. The slope of the voltage increase is, however, very similar to that for the 20-step voltage ramp with $F = 1000$. We observe that the data look very similar between the two ramps; however, the peaks at the first pulses after the pauses in the 20-step ramp are missing in the ramp with a single slope. We see again that the BD probability is higher at the voltages closer to the target value, but the overall result is the best for this run.

We also note here that the two-term exponential fits for the $\rho_{\mathrm{BD}}(n)$ support the previously proposed hypothesis of two mechanisms triggering a BD event; hence, the BDs can be classified as "primary", which are independent of other events and occur at a random place on the surface, and "secondary", which have an enhanced probability and are found to correlate stronger with the preceding BDs [31,37,40]. The best fits are with the scenarios with one slope and the one with 20 steps and $F = 4$ (coefficient of determination $R^2 = 0.87$ and $R^2 = 0.86$). However, we note that the behavior of the BDs during the ramping steps or slopes does not fit within this model, since they exhibit peaks related to the pauses, which affect the sBD probability stronger. Also, the BDs that occurred at early pulses with low voltages needed to be excluded from the fit, since they are strongly related to the level of surface contamination. It is also interesting to see that the ramping scenario with no ramping can be fitted with only one exponential term, suggesting that it has basically only secondary BDs—hence proving why the voltage ramping is necessary.

In the experiments with different pulsing repetition rates (see Fig. 8), we see a positive correlation of the BD probability with the pause length: The longer the pause between pulsing, the higher the breakdown probability within the initial pulses after the pause. Although differences between the pulsed dc systems and rf test stands make the direct comparison difficult, both of the results show the same trend. It should also be noted that each experiment demonstrates a difference in the time required for the onset of the effect, approximately 100 s for the dc case and $1 \times 10^5$ s for the rf case. These values correlate with the roughly estimated monolayer formation times, based on the impingement rate [41] of water molecules which are 80 and $6 \times 10^4$ s, respectively, at the internal pressure of each system. This supports the notion that some of the breakdowns may be triggered by vacuum residuals migrating to the high field regions of the surface during the idle time. The surfaces are consequently cleaned by the high-voltage pulses and breakdowns. Also, electrostatics of the surface impurities may play a role in the atom redistribution on the surface via surface migration processes [15].

We also observe a strong dependence of the BDR measured in bpp on the pulsing repetition rate, which was dramatically decreasing with the increase of the repetition rate. The trend is relatively smooth on the loglog scale over the whole repetition rate range, except for an unexpected data point, measured at a rate of 4 kHz. In Ref. [39], it was concluded that the repetition rate has only a negligible effect on the BDR; i.e., an increase in the repetition rate has no effect on the conditioning process. However, the conclusion was derived for the range of repetition rates from 25 to 200 Hz, while the wider range of repetition rates in the present study reveals the existence of such dependence. In Fig. 6, we see that the response over this narrower range is less significant, compared to the extremes of 10 Hz and 6 kHz.

Figure 6(b) shows that both the fraction of sBDs (out of all BDs) and the mean number of consecutive sBDs after a pBD decrease as the repetition rate increases; i.e., the idle time between the pulses decreases. This suggests that the idle time during pulsing affects especially the secondary BDs. We note that, again, the response is the greatest at the lowest and highest repetition rates. Between 50 and 1000 Hz, i.e., with idle times ranging from 20 to 1 ms, this trend is not visible.

The studies with the burst mode show that the idle time between the high-voltage pulses has also a significant effect on the BDR. In the last column in Table II, we show the ratio of the BDRs $R_{\mathrm{BDR}}$ during the odd- and even-numbered pulses in the burst mode with the frequencies of 100 Hz and 2 kHz. Although the BDR value obtained for the odd-numbered pulses was consistently larger than that measured for the even-numbered pulses, the difference is not dramatic (<2). It is clear that, although the difference in the pulse duration was more than 2 orders of magnitude, the BDR increases only insignificantly. Similar results were observed in the experiments with swapping of the repetition rates, shown in Fig. 7. The results of Ref. [39] and those shown in Table II and Fig. 7 indicate that the processes that take place during the idle time are slow and produce a more significant effect at longer idle times between the pulses. These results support the hypothesis of residual deposition from the vacuum. In other words, we clearly see that optimization of the surface conditioning procedure must include adjustment of the vacuum level and repetition rates of the high-voltage pulses.

## V. CONCLUSIONS

Several methods for investigating the effect of the various aspects of pulse timing on the breakdown rate were evaluated using the pulsed dc and rf systems. The studies include a comparison of different breakdown recovery scenarios, measurements of breakdown rates for variable repetition rates, and measurement of the effect of pauses between pulsing runs. All the measurements performed with pauses between pulsing ranging from 0.17 ms to 68 h show that the longer the idle time between pulses, the more prone the system is to breakdowns.

In comparison of different postbreakdown voltage recovery procedures, we observe a correlation between the increased idle time before pulsing, combined with a strong voltage increase, and the average number of secondary breakdowns, i.e., those that occur soon after and in the vicinity of the previous ones. This correlation suggests that the vacuum residuals must interact with the modification on the surface caused by the preceding breakdown, increasing the probability of a breakdown occurring during the postbreakdown recovery right after the pause and voltage increase.

The optimal voltage recovery after a breakdown was determined to be a linear increase in the voltage with the smallest idle time during the recovery. Thus, this ramping scenario will be used in future experiments with the pulsed dc systems. However, this result leaves room for future studies, especially focusing on the effect of the steepness of the ramping slope and smoothing the transition between the ramping and the target voltage pulsing.

In this work, we observe an enhanced breakdown rate following a period when the system was not pulsed. However, the enhancement is not dramatic, and the breakdown rate is restored rapidly after the system starts pulsing again, returning to the previous breakdown rate regardless of the length of the idle time.

## ACKNOWLEDGMENTS

The research and collaboration was made possible by the funding from the K-contract between Helsinki Institute of Physics and CERN. The authors also thank the Paul Scherrer Institute for funding the high-voltage generator that was supported by SNF/Funding LArge international REsearch projects Grant No. 20FL20 147463. We are grateful to the CLIC production team for design support on each stage of system development and sample fabrication.

## APPENDIX A: REPETITION RATE EXPERIMENTS IN DECREASING ORDER

As the measurements shown in Fig. 6 were conducted in increasing order of repetition rates, the presented results do not take into account the possibility of conditioning causing the downward trend in BDR (bpp) and upward trend in BDR (bps). To answer the concern, additional measurements were performed in decreasing order of repetition rates. The results are presented in Fig. 10.

The figure shows that, even when measured in the opposite order, the BDR measured in bpp still decreases as the repetition rate increases. Though less clear, there is also a slight increase in the BDR measured in bps. However, in both cases, the difference over the repetition rate range is smaller than when compared to the measurements performed in increasing order. This tells us that the conditioning might indeed play some role in the observed trend. Another thing to note is that the measurements in decreasing order covered a smaller range of repetition rates (3000–50 Hz) compared to the measurements in increasing order (10–6000 Hz).

## APPENDIX B: MORE STATISTICS ON PAUSE BETWEEN DC PULSING EXPERIMENTS

Figure 8(a) shows the probability of a BD occurring within the first second of running (2000 pulses) after a random pause between 6 and $10^5$ s. The graph shows the results of 939 measurements, averaged over eight logarithmically spaced bins. This figure shows an upward trend in the initial BD probability, though the uncertainty in the last three data points makes the result less reliable. This is mostly due to the fact that conducting as many measurements with long time intervals (>1000 s) as with short intervals is not feasibly possible.

In Fig. 11, we present the original data points and another way of analyzing the result to ensure that the drawn conclusion (the initial BD probability increasing as the pause time increases) is valid regardless of the large error estimates. In each measurement shown in the figure, a pulsing run with no ramping was performed after a pause of random length. The pulsing run ended in either a BD or when 50 000 pulses were reached. The initial BD probability is given as the fraction of runs that produced a BD in less than 2000 pulses and is shown as the moving average over 50 measurements.

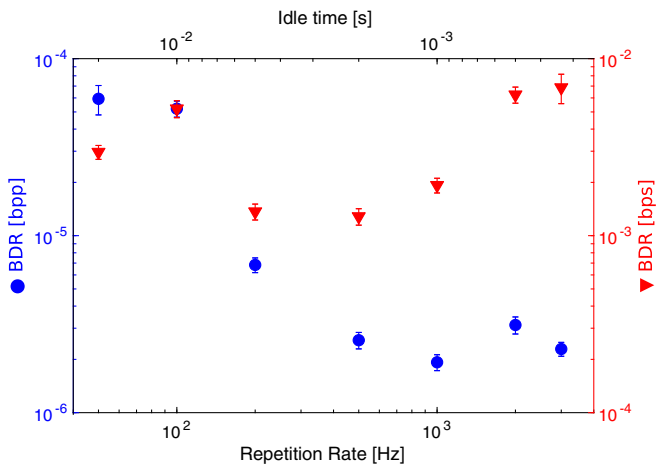

FIG. 10. BD experiments of Fig. 6(a) repeated, but now measured in decreasing order of repetition rates. The figure shows the BDR as BDs per pulse and as BDs per second for each repetition rate. The uncertainties were estimated using the standard error of the mean, though many of the error bars are too small to be visible.

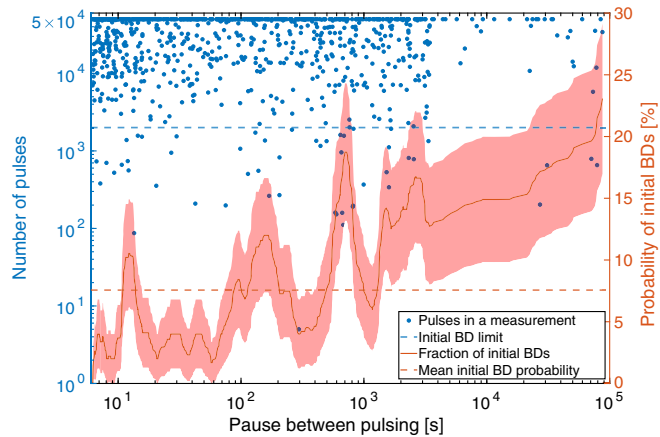

FIG. 11. Blue dots show the number of pulses in a pulsing run with no ramping, after a pause of random length. The solid orange line shows the moving average of the initial BD percentage [fraction of runs that resulted in a BD in less than 2000 pulses (blue dashed line)]. The red shaded area shows the uncertainty of the initial BD percentage as the standard error of the mean, and the orange dashed line shows its mean value.

The figure reinforces the initial conclusion that the initial BD probability indeed increases with an increasing pause in between the measurements. Also in this graph, the uncertainty is larger in the later data points, but we see that the increase is true even with these confidence limits.